# Approach to ergodicity in quantum wave functions

Bruno Eckhardt[1], Shmuel Fishman[2,3], Jonathan Keating[4],
Oded Agam[2], Jörg Main[5] and Kirsten Müller[6]

[1] FB Physik und Institut für Chemie und Biologie des Meeres,
C.v. Ossietzky Universität, Postfach 25 03, D-26111 Oldenburg, GERMANY

[2] Department of Physics, Technion, Haifa, 32000, ISRAEL

[3] Minerva Center for Nonlinear Physics of Complex Systems,
Technion, Haifa, 32000, ISRAEL

[4] Department of Mathematics, The University, Manchester, M13 9PL, UK

[5] Institut für Theoretische Physik, Ruhr-Universität Bochum, 44780 Bochum, GERMANY

[6] Centre de Recherches Nucléaires, Lab. Physique Theorique, F-67037 Strasbourg, FRANCE

According to theorems of Shnirelman and followers, in the semiclassical limit the quantum wave-functions of classically ergodic systems tend to the microcanonical density on the energy shell. We here develop a semiclassical theory that relates the rate of approach to the decay of certain classical fluctuations. For uniformly hyperbolic systems we find that the variance of the quantum matrix elements is proportional to the variance of the integral of the associated classical operator over trajectory segments of length $T_H$, and inversely proportional to $T_H^2$, where $T_H = h\bar{\rho}$ is the Heisenberg time, $\bar{\rho}$ being the mean density of states. Since for these systems the classical variance increases linearly with $T_H$, the variance of the matrix elements decays like $1/T_H$. For non-hyperbolic systems, like Hamiltonians with a mixed phase space and the stadium billiard, our results predict a slower decay due to sticking in marginally unstable regions. Numerical computations supporting these conclusions are presented for the bakers map and the hydrogen atom in a magnetic field.

## I. INTRODUCTION

In the semiclassical limit, quantum wavefunctions (or their corresponding phase space counterparts, such as the associated Wigner functions) are supported by classically invariant structures [1,2]. In integrable systems they are concentrated on tori [3,4] and in chaotic systems they tend to spread over connected components of the chaotic region [5]. The consequences of this for matrix elements have been analyzed by Shnirelman [6], Zelditch [7], Colin de Verdiere [8,9], and others [10,11]. Roughly speaking, their theorems state that in the semiclassical limit the matrix elements of smooth operators tend to the microcanonical average. Our aim here is to study the rate at which they do so.

The simplest way of defining the semiclassical limit is to allow a variable Planck's constant $h$ and to consider a sequence of values $h_n$ approaching zero such that for each $n$ some fixed energy $E_0$ is an eigenvalue of the Schrödinger equation. Then one has that for any smooth classical observable $A(\mathbf{p}, \mathbf{q})$ with which a reasonable quantum operator $\hat{A}$ can be associated, almost all diagonal matrix elements $\langle n|\hat{A}|n\rangle$ approach the classical microcanonical average

$$\langle A \rangle_{cl} = \int d\mathbf{p}\, d\mathbf{q}\, \delta(E_0 - H(\mathbf{p}, \mathbf{q})) A(\mathbf{p}, \mathbf{q})/\Omega \qquad (1)$$

as $n \to \infty$, where $\Omega$ is a normalization factor such that $\langle 1 \rangle_{cl} = 1$.

Since this definition of the semiclassical limit may be unfamiliar, and since it might on first sight appear artificial in a world in which $\hbar$ is in fact a constant, we point out two alternatives (see e.g. [12]). For the first, consider Hamiltonians homogeneous in positions and momenta, e.g. billiards, or systems with a suitable scaling of parameters, such as hydrogen in a magnetic field. Then it is possible to absorb Planck's constant into some power of the energy and so to map the semiclassical limit $h \to 0$ into the more familiar one of increasing energy or increasing quantum numbers.

The second alternative applies to general systems with non-scaling Hamiltonians. One then exploits the fact that the density of states near any energy $E_0$ will be semiclassically large, i.e., there are many states in an interval over which the classical mechanics does not change very much. Thus it is possible to expand the actions and other classical quantities to first order around the reference energy $E_0$. If the potential is bounded for all energies, the density of states will increase with increasing $E_0$, so that one can imagine covering the energy axis with intervals of fixed size which contain increasing numbers of states, but in which the classical mechanics is essentially fixed.

Common to all three approaches is the assumption that eigenstates can be labelled by integers $n$ which number either the values of the quantized Planck's constant, or the scaled energies, or the actual energies, in such a way that the semiclassical limit corresponds to $n \to \infty$.

The restriction on operators in the Shnirelman-type theorems is rather weak; it includes position and momentum operators, and smooth functions thereof, but it excludes projection operators since these do not have a smooth classical limit. More interesting is a restriction to 'almost all' eigenstates. This is quantified in terms of densities $d$ of subsets $\{E_{n_i}\}$ of states, defined as the quotient of the number of states in the set to the total number of states,

$$d = \lim_{n \to \infty} \frac{1}{n} \#\{n_i < n\}. \tag{2}$$

The Shnirelman-type theorems [6–10] hold for subsets of density one. Thus they still leave room for some individual wave functions to show relatively large deviations from the average and, perhaps, to be scarred in the neighborhood of short periodic orbits [13–15].

Several studies have verified that quantum matrix elements, both diagonal and off-diagonal, do indeed fluctuate around the classical limit [16–18]. Our concern here is with the variance of these fluctuations as the semiclassical limit is approached. Obviously when $\hbar$ vanishes the fluctuations must also vanish if Shnirelman's result is to be recovered. However, this decay may be rather slow, as mentioned by Colin de Verdière [9], and it may allow for large deviations due to scars in wavefunctions near periodic orbits [13,14]. In order to quantify these deviations we propose to look at the distribution of diagonal matrix elements,

$$P_{N,M}\,da = \mathrm{Prob}\left\{ \langle n|\hat{A}|n\rangle \in [a, a+da] \text{ and } N < n < N + M \right\} \tag{3}$$

as $N$ and $M$ both tend to infinity. Because of the scaling of the density of states in $d$-dimensional systems, one can take $M \sim N^\beta$ with $\beta < (d-1)/d$, so that the average includes an increasing number of states while the underlying classical mechanics is asymptotically fixed. The Shnirelman-type theorems then suggest that this distribution becomes narrower as one approaches the semiclassical limit $N \to \infty$.

Here we derive and test numerically a semiclassical expression for the variance of the matrix element distribution, relating it to the variance and a correlation function characterising certain classical fluctuations. Specifically, we focus our attention on the variance of matrix elements for states that are localized in one chaotic component. Using recent results from periodic orbit theory [19,12], we improve on the arguments of Feingold and Peres [16] and Prosen [18], who previously related fluctuations of matrix elements to classical phase space averages. Our analysis is somewhat similar to that of Wilkinson [20,21], but we are able to relax substantially his assumptions concerning the energy smoothing. We go beyond these studies by estimating the fluctuations of matrix elements around their classical averages and relating them directly to fluctuations of corresponding classical quantities. We ignore states localized in regular regions, but will consider the effects of classical sticking near islands in the chaotic sea. In particular, differences in the variances of matrix elements in hyperbolic and nonhyperbolic systems will be investigated.

The outline of the paper is as follows. In the next section we discuss the semiclassical derivation of the variance of matrix element distributions from several different points of view. Specifically, we describe an application of periodic orbit theory for hyperbolic systems, a reformulation of this approach in terms of a correlation function which is also be applicable to non-hyperbolic systems, and a derivation based on statistical properties of the classical motion. Some relations to random matrix theory and randomness assumptions for wave functions are discussed in section 3. In section 4, we present numerical data for the bakers map and for hydrogen in a magnetic field. We conclude in section 5 with a summary, some remarks on the stadium billiard, and additional general comments.

## II. SEMICLASSICAL MATRIX ELEMENT FLUCTUATIONS

In this section we relate fluctuations of diagonal matrix elements to properties of the periodic orbits and a correlation function of the corresponding classical motion. Our approach is explained in section II A, and for hyperbolic systems the semiclassical calculations are carried out in section II B using periodic orbit theory. Unfortunately, for nonhyperbolic systems we cannot use the resulting expression as it stands because of a lack of understanding of the role of the periodic orbits in this case. Instead, we derive a connection to a classical correlation function as an intermediate step within the original framework and then relate the fluctuations to an integral over this (section II C). We argue that the result can also be used in non-hyperbolic systems, as well as for chaotic components of mixed systems. The semiclassical matrix element distribution then depends on the decay of correlations in the classical system.

### A. Variances of matrix elements

On the quantum mechanical side, we consider the matrix element weighted density of states,

$$\rho^{(A)}(E) = \sum_n \langle n|\hat{A}|n\rangle \delta(E - E_n). \tag{4}$$

Without loss of generality, we assume that the operator $A$ has been shifted by its average, so that the matrix elements fluctuate around zero. Using a trick of Berry [22] the variance of the matrix elements, i.e. the average of their square, may be obtained from the square of the above density.

To avoid problems with the product of delta functions, we use smooth approximations, e.g. Gaussians of width $\epsilon$,

$$\delta_\epsilon(E) = \frac{1}{\sqrt{2\pi}\epsilon} e^{-E^2/2\epsilon^2}. \tag{5}$$

Crucially, the product of two Gaussians is again delta-function like, viz.,

$$\delta_\epsilon^2(E) = \frac{1}{2\sqrt{\pi}\epsilon}\delta_{\epsilon/\sqrt{2}}(E) \qquad (6)$$

The weighted density of states with these smoothed delta functions will be denoted by $\rho_\epsilon^{(A)}(E)$. Berry's approach then involves the square of $\rho_\epsilon^{(A)}$, multiplied by a factor proportional to the Gaussian smoothing parameter $\epsilon$,

$$K_\epsilon^{(A)}(E) = 2\sqrt{\pi}\epsilon \left(\rho_\epsilon^{(A)}(E)\right)^2 \qquad (7)$$

$$= \sum_n \sum_m \langle n|\hat{A}|n\rangle\langle m|\hat{A}|m\rangle 2\sqrt{\pi}\epsilon \cdot$$
$$\delta_\epsilon(E - E_n)\delta_\epsilon(E - E_m) . \qquad (8)$$

If we take $\epsilon$ smaller than the mean spacing between levels then, assuming there are no degeneracies, only the terms with $E_n = E_m$ contribute. Hence, using (6)

$$K_\epsilon^{(A)}(E) = \sum_n \left(\langle n|\hat{A}|n\rangle\right)^2 \delta_{\epsilon/\sqrt{2}}(E - E_n) \qquad (9)$$

from which we can estimate the variance by averaging over some energy interval $\Delta E$. Since the total number of states in such a range is $\bar{\rho}\Delta E$, $\bar{\rho}$ being the mean density of states (unweighted), the variance $\sigma_{\hat{A}}^2$ of the quantum matrix element distribution is given by

$$\sigma_{\hat{A}}^2(E) = \int_E^{E+\Delta E} K_\epsilon^{(A)}(E) \frac{dE}{\bar{\rho}\Delta E} . \qquad (10)$$

In the following subsections, we will derive various semiclassical and classical expressions for this fluctuation measure.

It should be noted that the choice of other functional forms for the smoothing of the delta functions changes some of the constants in the intermediate steps, but ultimately affects neither the final result for $K_\epsilon^{(A)}(E)$, nor the fact that this has a unique, well defined limit as $\epsilon \to 0$. A Gaussian smoothing has the advantage of being well localized in both the energy and time domains, so that there is little overlap between off-diagonal contributions to expressions like (8).

## B. Hyperbolic systems

We begin with the semiclassical approximation for $\rho^{(A)}(E)$ derived in [24],

$$\rho^{(A)}(E) = \sum_n \langle n|\hat{A}|n\rangle \, \delta(E - E_n)$$
$$= \int \frac{d\mathbf{p}\,d\mathbf{q}}{h^D} A(\mathbf{p},\mathbf{q})\delta(E - H(\mathbf{p},\mathbf{q})) +$$
$$+ \frac{1}{2\pi\hbar} \sum_p A_p w_p e^{iT_p E/\hbar} , \qquad (11)$$

where $p$ labels periodic orbits, of which the sum extends over positive and negative traversals, $D$ denotes the number of degrees of freedom,

$$A_p = \int_0^{T_p} A(\mathbf{p}(t),\mathbf{q}(t))\,dt \qquad (12)$$

is the integral of the observable along the $p$th orbit, and $T_p$ and $w_p$ are the orbit's period and weight. As explained in the introduction, we shall focus our attention on states in the neighborhood of $E_0$ and hence have expanded the orbit actions around this reference energy using $S_p(E') = S_p(E_0) + T_p(E_0)(E' - E_0)$, with $E = E' - E_0$, in which case the weights are given by

$$w_p = \frac{e^{iS_p(E_0)/\hbar - i\nu_p\pi/2}}{|\det(1 - M_p)|^{1/2}} \qquad (13)$$

where $M_p$ is the (monodromy) matrix of the linearization perpendicular to the orbit and $\nu_p$ is the Maslov index.

As before, we assume that the average of the operator vanishes so that the first term in (11) drops out. We then have to evaluate the square of the convolution of the semiclassical expression with the Gaussian smoothing (5). This gives as a semiclassical approximation to $K_\epsilon^{(A)}$

$$K_{\epsilon,sc}^{(A)}(E) = \frac{2\epsilon\sqrt{\pi}}{4\pi^2\hbar^2} \sum_{p'}\sum_{p''} A_{p'} A_{p''}^* w_{p'} w_{p''}^* \cdot$$
$$e^{i(T_{p'} - T_{p''})E/\hbar} e^{-\epsilon^2 T_{p'}^2/2\hbar^2} e^{-\epsilon^2 T_{p''}^2/2\hbar^2} , \qquad (14)$$

where the sum extends over all pairs of orbits and their negative traversals.

We now claim that the main contributions to this sum come from the diagonal terms for which $p' = p''$, so that

$$K_{\epsilon,sc}^{(A)}(E) \approx g\frac{2\epsilon\sqrt{\pi}}{2\pi^2\hbar^2} \sum_p |A_p|^2 |w_p|^2 e^{-\epsilon^2 T_p^2/\hbar^2} , \qquad (15)$$

where the $p$'s label individual orbits without negative traversals. If the system has time-reversal symmetry, then orbits come in pairs with the same phase and weight, giving rise to a symmetry factor $g = 2$. In systems without time-reversal symmetry, e.g. generic systems in a magnetic field, there is no such pairing and $g = 1$.

There are two ways to justify the neglect of the off-diagonal contributions. One source of cancellations results from the variation in the signs of the $A_p$'s, which must be present since the average $\langle A\rangle_{cl} = 0$. Assuming that the $A_p$'s are random, uncorrelated, and also have a vanishing mean one can justify (15) for any $\epsilon$ up to the limit set by the requirement that the Gaussians in (8) do not overlap, i.e. for $\epsilon < 1/\bar{\rho}$ (see section IID). It will be shown below that the diagonal approximation results in a well defined, $\epsilon$-independent value for the variance, despite this arbitrariness.

Another argument, which we will take advantage of in the non-hyperbolic case, appeals to Berry's analysis of spectral statistics [22]. If $A = 1$, then $A_p = T_p$ and (14) is directly related to the semiclassical expression for the form factor of the density of states. Classical and semiclassical sum rules as well as numerical observations [25] suggest that this form factor increases with increasing $t$ for $t < T_H$ and is constant for $t \gg T_H$, where $T_H = 2\pi\hbar\bar{\rho}$ is the Heisenberg time. As shown in [22], the range $t \ll T_H$ is well described by the diagonal contributions to (14). Since similar behaviour is also observed for matrix-element-weighted form factors [23], this implies that we can use the diagonal approximation up to times of the order $T_H$, and so suggests a critical size for $\epsilon$ of

$$\epsilon_c \approx \frac{1}{T_H} \approx \frac{1}{2\pi\hbar\rho} . \qquad (16)$$

This approximation is naturally expected to be better for the GUE than the GOE.

In the first argument, $\epsilon$ limits the periods of contributing orbits to $T_p < \hbar/\epsilon$, whereas in the second some $\epsilon_c$ is suggested which guarantees that the interferences between off-diagonal contributions are killed. In hyperbolic systems the choice of $\epsilon$ is not critical, but in nonhyperbolic systems with slowly decaying correlations $\epsilon$ determines a cut-off in the sum (15) which does affect the final expression. In such cases we will fix its value to be that given by (16), so that the sum is effectively truncated at the Heisenberg time $T_H$. Clearly, only the order of magnitude of $\epsilon$ is suggested by the above arguments, not the precise value, implying some variability in the semiclassical estimate for the matrix elements. One could improve on this if the correlations between actions and averages $A_p$ along periodic orbits were understood [25,26].

To evaluate (15) we have to appeal to some classical results [27–30]. From an analysis of the classical evolution operator one finds that the probability of returning to an infinitesimal tube around a periodic orbit $p$ after a time $T$ is given by

$$P_p(T) = \delta(T_p - T)\frac{T_p}{|\det(M_p - 1)|} \qquad (17)$$

$$= \delta(T_p - T)\,T_p\,|w_p|^2 . \qquad (18)$$

When summed over all orbits with periods $T_p \in [T, T + \Delta T]$ for sufficiently large $T$, this satisfies

$$\int_T^{T+\Delta T}\sum_p P_p(T)dT = \Delta T , \qquad (19)$$

since the periodic orbits approximate the invariant measure. The density of orbits increases like $e^{h_t T}/T$, $h_t$ being the topological entropy, so that the weights $|w_p|^2$ have to decrease on average like $e^{-h_t T}$. The integrals $A_p$ for orbits in this interval will fluctuate around zero. Assuming that correlations along the orbits decay sufficiently rapidly, as in hyperbolic systems, the contributions to the integration of the observable along the orbit will fluctuate randomly between positive and negative values, so that the distribution of $A_p$'s for orbits with periods near $T$ will be Gaussian with a variance that increases linearly with $T$,

$$P_T(a)da = \text{Prob}\{A_p \in [a, a + da], T_p \text{ near } T\}$$
$$= \frac{1}{\sqrt{2\pi\alpha T}}e^{\frac{-a^2}{2\alpha T}}da . \qquad (20)$$

Combining the sum rule (19) with the variance implied by (20) gives

$$T\sum_{T < T_p < T + \Delta T}|A_p|^2|w_p|^2 = \alpha T\,\Delta T \qquad (21)$$

so that upon replacing the sum over orbits in the diagonal approximation (15) by an integral, we find

$$K_\epsilon^{(A)}(E) = g\frac{2\epsilon\sqrt{\pi}}{2\pi^2\hbar^2}\int_0^\infty dT\,\alpha e^{-\epsilon^2 T^2/\hbar^2} \qquad (22)$$

$$= g\frac{\alpha}{\hbar} , \qquad (23)$$

where $g$ is again the symmetry factor. With the choice (16) for $\epsilon$, the integral is effectively truncated at the Heisenberg time. However, because of the distribution (20), the final result is actually independent of $\epsilon$, just as the quantum expression itself is. The semiclassical estimate of the variance of the matrix elements follows from this after averaging over some energy interval, as explained before (10). The result, which is the main one of this section, is that

$$\sigma_{\hat{A},sc}^2 = g\frac{\alpha}{T_H} . \qquad (24)$$

This shows first of all that the variance decays as the inverse of the Heisenberg time, and secondly, since $\alpha$ is determined from the classical trajectories, that classical and quantum fluctuations are related.

An alternative way of writing this relation is to consider not the distribution of the integral of the observable along the orbits, the $A_p$'s, but the distribution of the averages $a_p = A_p/T_p$. This is again a Gaussian, but now with a variance $\sigma_a^2(T) = \alpha/T$. The implication is then that up to a symmetry related factor the widths of the quantum matrix element distribution and the distribution of classical averages from trajectory segments of length $T_H$ are the same. And since the latter decays like $1/T_H$, the quantum matrix elements narrow around the classical average at the same rate.

## C. Correlation functions and nonhyperbolic systems

The situation is more complicated if a system is not nicely hyperbolic but rather has a mixed phase space or marginally stable orbits. Both around islands (due to trapping in the nested cantorus structure) and near marginally stable orbits (because of the slow escape) one finds increased staying times, resulting in more slowly decaying correlations and possible non-Gaussian distributions [31,32] with more slowly decaying variances. Furthermore, in the case of marginally stable orbits, the semiclassical weights diverge and the Gutzwiller trace formula used above has to be improved [33–35]. To investigate the behavior of the connected chaotic component in such a system it will be useful to obtain an expression for the variance of matrix elements that does not depend explicitly on the weights of periodic orbits in any non-hyperbolic regions. We now derive such an expression in terms of a classical auto-correlation function.

We begin by relating $A_p^2$ to a correlation function along periodic orbits. Abbreviating the phase space argument of the classical observable $A(\mathbf{p}(t), \mathbf{q}(t))$ by $\mathbf{z}(t)$, with an index indicating the periodic orbit, we have that

$$
\begin{aligned}
A_p^2 &= \int_0^{T_p} dt_1 \int_0^{T_p} dt_2 \, A(\mathbf{z}_p(t_1)) A(\mathbf{z}_p(t_2)) \\
&= \int_0^{T_p} d\tau' \int_{-T_p/2}^{T_p/2} d\tau \, A(\mathbf{z}_p(\tau'+\tau/2)) A(\mathbf{z}_p(\tau'-\tau/2)) \\
&= T_p \int_{-T_p/2}^{T_p/2} d\tau \, C_p(\tau)
\end{aligned}
\tag{25}
$$

where periodicity of the integrand has been exploited and

$$
C_p(\tau) = \frac{1}{T_p} \int_0^{T_p} d\tau' \, A(\mathbf{z}_p(\tau'+\tau/2)) A(\mathbf{z}_p(\tau'-\tau/2))
\tag{26}
$$

is the auto-correlation function along the periodic orbit. When substituted in the orbit sum (15) one can again use the fact that the weighted periodic orbits approximate the invariant density so that the average of $C_p(\tau)$ over all orbits becomes the classical correlation function (for more details and a quantitative comparison in the hyperbolic case, see [36]). Thus we can write, in analogy to (21),

$$
\begin{aligned}
\sum_{T<T_p<T+\Delta T} |A_p|^2 |w_p|^2 &= \int_{-T/2}^{T/2} d\tau \sum_{T<T_p<T+\Delta T} C_p(\tau) \, T_p |w_p|^2 \\
&\approx \int_{-T/2}^{T/2} d\tau \, C(\tau) \, \Delta T
\end{aligned}
\tag{27}
$$

where

$$
C(\tau) = \langle A(\mathbf{z}(0)) A(\mathbf{z}(\tau)) \rangle
\tag{28}
$$

is the average correlation function as determined from non-periodic ergodic trajectories or, equivalently, by averaging over the invariant measure.

The semiclassical expression (15) for the variance may thus be written

$$
\begin{aligned}
K_{\epsilon,sc}^{(A)}(E) &= g \frac{2\epsilon\sqrt{\pi}}{2\pi^2\hbar^2} \int_0^\infty dT \int_{-T/2}^{T/2} d\tau \, C(\tau) e^{-\epsilon^2 T^2/\hbar^2} \\
&= g \frac{1}{\pi\hbar} \int_0^\infty d\tau \, C(\tau) f_\epsilon(\tau) \,,
\end{aligned}
\tag{29}
$$

where

$$
f_\epsilon(\tau) = 1 - \mathrm{erf}(2\tau\epsilon/\hbar)
\tag{30}
$$

with $\mathrm{erf}(x) = \frac{2}{\sqrt{\pi}} \int_0^x \exp(-z^2) dz$. The precise form of $f_\epsilon$ reflects the fact that we used a Gaussian smoothing; had we worked instead with Lorentzians, we would have obtained $f_\epsilon(\tau) = \exp(-2\tau\epsilon/\hbar)$, and hence a Laplace Transform in (29). The above expression, together with the choice (16) for $\epsilon$ seems to be as far as one can generally go in non-hyperbolic cases. It relates the quantum fluctuations to an integral of a classical correlation function.

The results of the previous section can be recovered if the correlations decay sufficiently rapidly. Then the integral over the correlation function tends to a constant

$$
\eta = \lim_{T\to\infty} \int_{-T}^{T} d\tau \, C(\tau) = 2 \int_0^\infty d\tau \, C(\tau) \,,
\tag{31}
$$

and so in the limit $\epsilon \to 0$, in which $f_\epsilon \to 1$, we arrive at the estimate

$$
K_\epsilon^{(A)}(E) \approx g\frac{\eta}{\hbar} \,.
\tag{32}
$$

To connect this $\eta$ (eq. 31) with the variance $\alpha T$ of the $A_p$'s (20), note that for a system with correlations decaying exponentially on a time scale $\lambda$, one can write

$$
C(\tau) = \frac{\alpha\lambda}{2} e^{-\lambda\tau} \,.
\tag{33}
$$

Substitution into (31) thus gives $\eta = \alpha$. If the correlations decay more slowly, one expects $\eta > \alpha$.

The relation between the variance of the matrix element distribution and the classical correlation function is now also applicable in cases of slowly decaying correlations and nonhyperbolic systems, since the decay of correlations is directly related to trapping in regions in which the periodic orbit representation of the invariant measure becomes suspect. If the long time behavior is dominated by regions where the probability to be trapped for a time $\tau$ decays as $\tau^{-\gamma}$, then the correlation function also decays in the same way:

$$
C(\tau) \sim \tau^{-\gamma} \,.
\tag{34}
$$

This results from the fact that $A$ takes similar values (different from the vanishing mean) in the region where the motion is trapped for a long time.

The above ansatz (34) for the asymptotic behaviour of the correlation function is integrable for exponents $\gamma > 1$, leading to the same scaling for the variance of the matrix elements as in the hyperbolic case. For $\gamma \le 1$, the integral diverges and so is dominated by the effective cut-off at the Heisenberg time. Consequently, one has for the asymptotic behaviour of the semiclassical variance

$$
\sigma_{\tilde{A},sc}^2 \sim \begin{cases} T_H^{-1} & \text{for } \gamma > 1 \\ \ln T_H/T_H & \text{for } \gamma = 1 \\ T_H^{-\gamma} & \text{for } \gamma < 1 \end{cases}
\tag{35}
$$

The decay of correlations has been studied for a variety of chaotic systems. In particular, for trapping near elliptic islands embedded in a chaotic sea the correlations have been found to decay agebraically, with the exponent $\gamma$ in the range $\gamma \sim 1.2 - 1.5$ [37–41]. This decay is determined by the winding numbers and cantori of the surrounding islands and thus may be visible only after exceedingly long times [41,42]. The stadium billiard falls into the middle category, since there $\gamma = 1$ [43].

### D. A derivation based on classical randomness

For hyperbolic systems, correlations between trajectories decay sufficiently fast that for long orbits different $A_p$ can be considered independent random variables, leading to

$$
\begin{aligned}
\langle\!\langle A_{p'} A_{p''} \rangle\!\rangle &= \int_0^{T_{p'}} d\tau_1 \int_0^{T_{p''}} d\tau_2 \, \langle A(\mathbf{z}_{p'}(\tau_1)) A(\mathbf{z}_{p''}(\tau_2)) \rangle \\
&= \langle A_p^2 \rangle \, \delta_{p',p''} = \alpha T \, \delta_{p',p''} \,,
\end{aligned}
\tag{36}
$$

where $\langle\!\langle \cdots \rangle\!\rangle$ is an average over all pairs of periodic orbits of period $T_p \in [T, T+dT]$ and $\langle A_p^2 \rangle$ is the variance of the periodic orbit integrals as calculated from the distribution (20). This result should also hold if the trapping is not too strong, since then the times at which different orbits enter the trap are uncorrelated. Averaging (14) over periodic orbits in a small interval of periods in the vicinity of $T$, perhaps supplemented by an average over a small energy interval as well, leads to (15). Importantly, under these assumptions the diagonal approximation is not limited to orbits of period shorter than the Heisenberg time. (It is worth pointing out that the difference between (14) and the corresponding expression for the spectral form factor is that the periodic orbit contributions are in this case proportional to $A_p$, and the additional randomization thus introduced is enough to kill the off-diagonal terms for all sufficiently large $T$.) The calculation leading from (15) to (24) can then be repeated without the restriction on the periods of the orbits, resulting in the semiclassical matrix element variance

$$
\sigma_{A,sc}^2 = g\frac{1}{\bar{\rho}}\frac{\epsilon\sqrt{\pi}}{\pi^2\hbar^2}\int_0^\infty dt \, \frac{\langle A_p^2 \rangle(T)}{T} \, e^{-\epsilon^2 T^2/\hbar^2}.
\tag{37}
$$

If the correlations decay sufficiently rapidly, the distribution of the $A_p$'s is Gaussian and given by (20), so that we again arrive at

$$
\hat{\sigma}_A^2 = g\frac{\eta}{2\pi\hbar\bar{\rho}} = g\frac{\eta}{T_H}
\tag{38}
$$

in agreement with the estimate (32).

### III. RESULTS OF THEORIES ASSUMING RANDOMNESS

In this section the predictions for the variance of the diagonal matrix elements in the framework of theories that assume true randomness will be summarized. These will be compared with the predictions of periodic orbit theory derived in Sect. 2.

### A. Random Matrix Theory

There is much evidence that many quantum properties of chaotic systems are consistent with random matrix theory [44], although there is no rigorous proof for this. It is therefore of interest to relate our predictions to the corresponding random matrix results. For a typical observable $A$, random matrix theory predicts [45,46] that

$$
\hat{\sigma}_A^2 = g\hat{\sigma}_{A,off}^2 \,,
\tag{39}
$$

where $\hat{\sigma}_{A,off}^2$ is the variance of the off-diagonal matrix elements, while $g$ depends on the symmetry of the model and takes the values 1 and 2 for the GUE and GOE respectively. In the semiclassical limit the variance of the off-diagonal matrix elements is related to the classical correlation function by [16],

$$
\hat{\sigma}_{A,off}^2 = \frac{\eta}{T_H} = \frac{2}{T_H}\int_0^\infty dt\, C(t)\,.
\tag{40}
$$

Therefore, eqs. (39) and (40) agree with eqs. (32) and (24) which were obtained directly from periodic orbit theory under the assumption that the classical correlations decay sufficiently rapidly.

### B. Random Wave Function Theory

This theory assumes that eigenstates of chaotic systems are Gaussian random functions [47,48]. In particular, if we concentrate on two dimensional billiards, the wave function at each point $\mathbf{q}$ of the coordinate space satisfies the normalization,

$$
\langle \psi^2(\mathbf{q}) \rangle_{st} = \frac{1}{\Omega_r}
\tag{41}
$$

where $\Omega_r$ is the area of the billiard. It is assumed that the system is invariant under time reversal and therefore the wave function can be considered real. Here $\langle\cdots\rangle_{st}$ denotes a statistical average over an ensemble of similar billiards, or over a small region in space around $\mathbf{q}$. The correlation function between the wave function evaluated at different points is [47,48],

$$
\langle \psi(\mathbf{q}_1)\psi(\mathbf{q}_2) \rangle_{st} = \frac{1}{\Omega_r}J_0\left(k|\mathbf{q}_1 - \mathbf{q}_2|\right)
\tag{42}
$$

where $J_0$ is a Bessel function and $k$ is the wave number. The statistical average of a diagonal matrix element reduces to the microcanonical average of the observable, namely

$$\langle\langle\psi|\hat{A}|\psi\rangle\rangle_{st} = \langle A\rangle_{cl} \qquad (43)$$

which is set to vanish in the present paper. The variance is

$$\langle\langle\psi|\hat{A}|\psi\rangle^2\rangle_{st} = \int\int d\mathbf{q}_1 d\mathbf{q}_2 \langle\psi^2(\mathbf{q}_1)\psi^2(\mathbf{q}_2)\rangle_{st} A(\mathbf{q}_1) A(\mathbf{q}_2), \qquad (44)$$

where we restrict ourselves to observables that depend only on coordinates. For Gaussian random functions satisfying (41) and (42),

$$\langle\psi^2(\mathbf{q}_1)\psi^2(\mathbf{q}_2)\rangle_{st} = \frac{1}{\Omega_r^2}\left(2J_0^2(k|\mathbf{q}_1 - \mathbf{q}_2|) + 1\right) \qquad (45)$$

leading to the corresponding result for the variance of the matrix element distribution,

$$\hat{\sigma}_{A,r}^2 = \langle\langle\psi|\hat{A}|\psi\rangle^2\rangle_{st}$$
$$= \frac{2}{\Omega_r^2}\int\int d\mathbf{q}_1 d\mathbf{q}_2 J_0^2(k|\mathbf{q}_1 - \mathbf{q}_2|) A(\mathbf{q}_1) A(\mathbf{q}_2)$$

In order to obtain a semiclassical expression, we use the (short wavelength) asymptotic form of the Bessel function $J_0(z) \sim \sqrt{\frac{2}{\pi z}}\cos(z - \pi/4)$ and approximate $\cos^2(z - \pi/4)$ by its averaged value $\frac{1}{2}$. The result is that

$$\hat{\sigma}_{A,r}^2 = \frac{2}{\Omega_r^2 \pi k}\int\int d\mathbf{q}_1 d\boldsymbol{\Delta}\mathbf{q}\, \frac{A(\mathbf{q}_1) A(\mathbf{q}_1 + \boldsymbol{\Delta}\mathbf{q})}{|\boldsymbol{\Delta}\mathbf{q}|} \qquad (46)$$

where $\boldsymbol{\Delta}\mathbf{q} = \mathbf{q}_2 - \mathbf{q}_1$. This can also be written as

$$\hat{\sigma}_{A,r}^2 = \frac{2}{\pi L_H}\frac{1}{\Omega_r}\int\int d\mathbf{q}_1 d\boldsymbol{\Delta}\mathbf{q}\, \frac{A(\mathbf{q}_1) A(\mathbf{q}_1 + \boldsymbol{\Delta}\mathbf{q})}{|\boldsymbol{\Delta}\mathbf{q}|}, \qquad (47)$$

where $L_H = T_H v$ is the Heisenberg length, $v$ being the speed. It clearly scales with the Heisenberg time in accordance with (24) and (32). Since one can obtain (44) under the same semiclassical assumptions as enter our derivation, this result agrees with the previous expression (O. Agam, unpublished).

## IV. NUMERICAL EXAMPLES

Before turning to our numerical tests, we explain a useful classical approximation to the sum over periodic orbits in (15). Since in hyperbolic systems the invariant density in phase space can be approximated by $\delta$-functions on the periodic points with weights given by $|w_p|^2 T_p$, the above expression can, by ergodicity, also be computed by following a nonperiodic trajectory, which is then subdivided into segments of length $T$. This has been exploited in the numerical calculations below.

### A. Bakers map

As an example of a nicely hyperbolic system we take the bakers map, with the quantization proposed by Saraceno [49–51]. In the form given, the above formulae do not strictly apply to maps. Nevertheless, the time evolution is represented by a unitary operator, the traces of powers of which have a semiclassical periodic orbit expansion [52–55]. In the case of the bakers map, where the classical phase space is bounded, the unitary matrix has dimension $N$, which corresponds to the inverse of Plancks constant, and all its properties can be obtained from traces of the first $N$ powers. Hence, in the semiclassical expression (24), we should allow for all orbits up to period $N$.

Because of a rapid decay of correlations in the bakers map, some form of the law of large numbers applies to averages of observables over several time steps. The variance of the observable summed over $n$ time steps is given by $\langle A_p^2\rangle(n) = \alpha n$. Moreover, the distribution function is, to a good approximation, Gaussian, as in (20). In particular, for the observable $A(p, x) = \cos(2\pi x)$ one has $\alpha = 1/2$.

Quantum eigenstates and eigenvectors were obtained by direct diagonalization of the unitary propagator. To improve on the statistics of the matrix elements, small groups of matrices have been collected together. If $N = 100, 200$ and $400$ denotes the central value, the included matrices are of size $N$, $N\pm 2$ and $N\pm 4$. When rescaled by $\sqrt{N}$, the quantum matrix elements for $\cos(2\pi x)$ also follow a Gaussian distribution, as shown in Fig. 2. Quantitatively, the variance of their distribution decreases with $N$ according to

$$\sigma_{\hat{A},qm}^2(N) \approx \tilde{\alpha}/N, \qquad (48)$$

where $\tilde{\alpha}$ is about $1.5\ldots1.8$, the larger value corresponding to the largest $N$.

The relation between the classical and quantum scaling with $N$ is given by $\tilde{\alpha} = 4\alpha$: one factor of two coming from the fact that the quantum map has an antiunitary symmetry and so $g = 2$, and the second factor of two because it also has a unitary symmetry and the two corresponding symmetry-classes contribute independently. The difference of about 10% can probably be traced back to the large corrections to the semiclassical approximation in the bakers map due to diffraction from the discontinuity [55,56].

### B. Hydrogen in a magnetic field

As a second example we consider a smooth dynamical system: the hydrogen atom in a strong magnetic field (for reviews, see [57,58]). At the energy to be considered here, the ergodic component covers a large fraction of phase space, so that no elliptic islands are visible in a surface of section. The tiny islands present are rarely visited and only influence the very long time behaviour. They will not be noticable in our calculations where the Heisenberg time is small and fluctuations of short trajectory segments dominate. Nevertheless, the motion of orbits moving close to the field axis is approximately separable and thus correlations involving these orbits may decay rather weakly.

Both the classical and quantum system have scaling properties which ease the calculations, but which may obscure the relation to the previous analysis. We thus provide some details of the transformations involved. After separation of the trivial degree of freedom - a rotation around the magnetic field axis - the classical Hamiltonian for the quadratic Zeeman effect of a hydrogen atom in a strong magnetic field at zero angular momentum around the field axis becomes

$$H = \frac{p_\rho^2 + p_z^2}{2m} - \frac{q^2}{4\pi\varepsilon_0}\frac{1}{\sqrt{\rho^2+z^2}} + \frac{q^2B^2}{8m}\rho^2 \ , \quad (49)$$

where $(z, \rho)$ are the coordinates parallel and perpendicular to the field, $(p_z, p_\rho)$ are the conjugate momenta, $q$ is the charge of the electron, and $B$ the magnetic field. Rescaling to atomic units introduces some hidden $\hbar$ dependencies. With the Bohr length $a_0 = 4\pi\varepsilon_0\hbar^2/q^2 m$, the unit of momentum $p_0 = \hbar/a_0$, the energy $E_0 = \hbar^2/ma_0^2$ and the unit for the magnetic field $B_0 = mE_0/q\hbar$, one finds the rescaled Hamiltonian (all variables denoted by a prime)

$$H' = H/E_0 = \frac{p_\rho'^2 + p_z'^2}{2} - \frac{1}{\sqrt{\rho'^2+z'^2}} + \frac{\gamma^2}{8}\rho'^2 \ , \quad (50)$$

where $\gamma = B/B_0$ is the dimensionless magnetic field. The $\gamma$-dependence on the right hand side can be eliminated by the rescaling

$$p' = \gamma^{1/3}p'' \qquad , \qquad r' = \gamma^{-2/3}r'' \quad (51)$$

and similarly for $\rho'$, whereby

$$H'' = \gamma^{-2/3}H/E_0 = \frac{p_\rho''^2 + p_z''^2}{2} - \frac{1}{\sqrt{\rho''^2+z''^2}} + \frac{1}{8}\rho''^2 \ . \quad (52)$$

This shows that the classical dynamics depends only on the rescaled energy $H'' = \varepsilon = \gamma^{-2/3}E/E_0$. Note that this quantity is independent of $\hbar$. The relation between the original time $t$ and the dimensionless and rescaled time $t''$ in this latter system is given by

$$t = \frac{\hbar}{E_0}\gamma^{-1}t'' \ . \quad (53)$$

Upon quantization, eq. (52) represents a generalized eigenvalue problem for $\gamma^{1/3}$ if $\varepsilon$ is held constant, i.e.

$$-(\gamma^{1/3})^2\frac{1}{2}\Delta - \frac{1}{\sqrt{\rho''^2+z''^2}} + \frac{1}{8}\rho''^2 = \varepsilon \ . \quad (54)$$

Clearly one may interpret $\gamma^{1/3}$ as an effective Plancks constant and study the semiclassical limit (as $\gamma^{1/3} \to 0$) of matrix elements with the classical mechanics fixed. (It is worth pointing out that $\gamma^{1/3}$ approaches zero as the energy increases up to the ionization limit $E = 0$ or, equivalently, as the eigenvalue $\gamma^{-1/3} = \sqrt{\varepsilon E_0/E}$ goes to infinity). This is obviously an example of the procedure described in the Introduction, in that one is quantizing (the effective) Planck's constant ($\gamma^{1/3}$) so that the fixed rescaled energy $\varepsilon$ is an eigenvalue of (54).

The integrated density of states in this system is given by

$$N(E, B, \hbar) = \int \frac{dp_\rho\, dp_z\, d\rho\, dz}{(2\pi\hbar)^2}\Theta(E - H)$$
$$= \gamma^{-2/3}\Omega(\varepsilon) \quad (55)$$

with

$$\Omega(\varepsilon) = \int \frac{dp_\rho''\, dp_z''\, d\rho''\, dz''}{(2\pi)^2}\, \Theta(\varepsilon - H''(p_\rho'', p_z'', \rho'', z'')) \quad (56)$$

and thus depends on energy, magnetic field and Planck's constant only through the dimensionless combinations $\gamma$ and $\varepsilon$. From this expression one can calculate the mean density of states as a function of energy at fixed magnetic field and Planck's constant,

$$\overline{\rho(E)} = \frac{\partial N}{\partial E} = \frac{1}{E_0}\gamma^{-4/3}\frac{\partial\Omega}{\partial\varepsilon} \ . \quad (57)$$

This gives the Heisenberg time $T_H$ in original coordinates

$$T_H = 2\pi\hbar\overline{\rho(E)} = 2\pi\gamma^{-4/3}\frac{\partial\Omega}{\partial\varepsilon}\frac{\hbar}{E_0} \quad (58)$$

and thus, by (53),

$$T_H'' = 2\pi\gamma^{-1/3}\frac{\partial\Omega}{\partial\varepsilon} \quad (59)$$

in rescaled, dimensionless coordinates.

For reasons of numerical convenience we choose to test the present theory for $\hat{A} = 1/2r$, which is smooth enough to allow for the application of semiclassical approximations [24]. According to (51), the average of this variable over the energy shell at fixed energy and magnetic field scales like $\gamma^{2/3}$. Taking this scaling out, one finds a stationary average for the matrix elements

$$A_n = \gamma_n^{-2/3}\langle n|\hat{A}|n\rangle \ . \quad (60)$$

The classical average of the Weyl symbol of $\hat{A}$ is given by an integral like that for the density of states and can be evaluted numerically to be

$$A_{cl} = 0.259 \ . \quad (61)$$

The quantum analysis is based on the calculation of the first 9750 positive $z$-parity states for $\varepsilon = -0.1$. These were obtained using the Lanczos algorithm to diagonalize the strongly banded Hamiltonian in a harmonic oscillator basis. The scaled matrix elements vs. the quantized values of $\gamma^{-1/3}$ are shown in Fig. 3a. The histogram, Fig. 3b, over all the states shows an almost Gaussian distribution around classical microcanonical average.

Fig. 4 compares classical and quantum variances as a function of $T_H$. The classical calculation is based on trajectory segments of length $T_H$. The quantum data are obtained by a Gaussian smoothing of width $\epsilon = 5$ in $K_\epsilon^{(A)}(\gamma^{-1/3})$ (cf. eqn. 9). The classical variances decay slower than $1/T_H$ (with an exponent of about $-1/2$), presumably due to a slower decorrelation of motion parallel to the field. Thus eqns (34) and (35) have to be applied and the exponents found are in accord with the prediction. As far as the constants are concerned, the agreement is unexpectedly good since in this case the prefactors actually depend on the choice of the (arbitrary) smoothing parameter $\epsilon$.

## V. SUMMARY AND DISCUSSION

The main result presented here is a connection between the quantum fluctuations of matrix elements around the classical microcanonical average and fluctuations of classical averages over orbit segments of length of the order of the Heisenberg time, as given by (24) in the hyperbolic case and (29) and (10) in the nonhyperbolic case. These results also establish the relation (39) between diagonal and off-diagonal matrix elements within periodic orbit theory, a direct link elusive to the approach of Peres, Feingold and Wilkinson [16,20]. In the case of the bakers map there is acceptable agreement between quantum behaviour and the semiclassical predictions, and in the case of the hydrogen atom in a strong magnetic field it is fair. The extensions proposed to non-hyperbolic systems could not be tested in depth here for a lack of a sufficient number of eigenstates. The evidence from the hydrogen example (where the correlations did not decay very rapidly) suggests that the relation persists. Further tests should perhaps be based on maps since there one can probe the semiclassical limit much more deeply than with smooth systems.

An investigation of non-hyperbolic systems should also include a discussion of the behaviour of the full distribution of matrix elements and classical short time averages, rather than just the variance. The comparison of classical and quantum variances for the quadratic Zeeman effect shows that even in non-ideal cases the two are of the same size. If there are islands of quasi-integrable motion, the distribution of matrix elements has a wing dominated by states localized on and near them. It is known that in such cases the classical distribution develops algebraic tails [31,32]. Our statistics are not good enough to say much quantitatively about this relation. It would be worthwhile to see how far the connection between the above comments about trapping in and near to the bouncing ball mode, or in systems with cantori, and fluctuations in the matrix elements can be carried. This may also be of interest in mesoscopic systems, since one might speculate that a nice hyperbolic classical system will show Gaussian fluctuations very much as in a disordered metal. A system with mixed phase space will show different fluctuations and one might ask whether they are the same as those of a strongly disordered system showing localization.

Along similar lines, an investigation of matrix elements between high lying states of the stadium billiard might be worthwhile, since there one expects on the classical side large effects due to the bouncing ball modes [59] and the slow decay of correlations [43], while the quantum side is strongly influenced by scars [13,14].

Extensive work, in particular on the kicked rotator [60,61], has shown that many states are localized and change the distribution of matrix elements. This effect is clearly not captured by our result, although it is not obvious where the derivation has to be modified. Because of their practical importance, for instance in low frequency AC conductivity, as well as noise induced diffusion (see [62,63] and references therein), this question deserves further attention.

The results presented here can also be used to estimate quantitatively the error committed by a classical calculation, say for excitations of a molecule [64] or for conductivity [65]. In both cases the quantum expression for the correlation function will fluctuate around the appropriate value of the classical correlation function, with a sigma variation given by the classical statistical variation of trajectories up to the Heisenberg time.

All of the calculations presented in this paper were carried out within the framework of the diagonal approximation of the sum (15). This ignores the behaviour in the short-time regime, where individual orbits are important. Specifically, our calculations apply to the asymptotic regime in which the Heisenberg time $T_H$, the time scale relevant for the results (32), (35) and (35), is much larger than the periods of the shortest periodic orbits. There is, of course, a wide range of energies where the short orbits may be of interest, as demonstrated by the existence of scars [13,14]. They also influence thermodynamic properties at intermediate temperatures [66]. The exact contribution of these to the fluctuations of matrix elements is left for further study.

## ACKNOWLEDGMENTS

It is our pleasure to thank L. Bunimovich and M. Wilkinson for informative discussions. Two of us (BE and JK) wish to thank the Institute for Theoretical Physics at the Technion for support and the stimulating environment in which this work could prosper. The work was further supported by the Minerva Center for Non-linear Physics of Complex Systems at the Technion and the Weizmann Institute (BE), the Deutsche Forschungsgemeinschaft (KM) and the US-Israel Binational Science Foundation and the Fund for the Encouragement of reserach at the Technion. We are grateful for all this support.

---

FIG. 1.

Scaled distributions of classical observables in the bakers map. The distributions are for $n = 50$, 100, 200, 400 and 800 iterations of the map as computed from a long ergodic trajectory. The number of segments used to obtain the distribution is 16000 for the shortest and 1000 for the largest iteration number.

FIG. 2.

Distribution of the matrix elements for the observable $\hat{A} = \cos 2\pi x$ in the quantized bakers map. Five sets of matrix elements of size $N$, $N \pm 2$ and $N \pm 4$ have been superimposed and the central $N$ values are listed. All matrix elements have been rescaled by $\sqrt{N}$ and the histograms have been superimposed. The Gaussian has a width of about 1.8, close to the estimated value of $\hat{\alpha}$.

FIG. 3.

Scaled expectation values of $\hat{A} = 1/2r$ vs eigenvalue $z_n = \gamma^{-1/3}$ for the lowest 9750 eigenstates in the positive parity subspace of the quadratic Zeemann effect at $\epsilon = -0.1$ (a). The histogram in (b) shows the normalized distribution of matrix elements. Note the clustering around the classical average of 0.259 and the almost Gaussian form.

FIG. 4.

Variance of quantum matrix elements and averages over classical trajectory segments vs. Heisenberg time $T_H/2\pi$. The full line for the quantum matrix elements has been obtained with a Gaussian smoothing of width $\epsilon = 5$ in eq. (9). The dashed curve was calculated from averages over classical trajectory segments of length $T$. The inset shows the same data on a log-log scale and reveals a transition from a $T^{-1/2}$ dependence for short times to the anticipated hyperbolic $T^{-1}$ law for larger times.

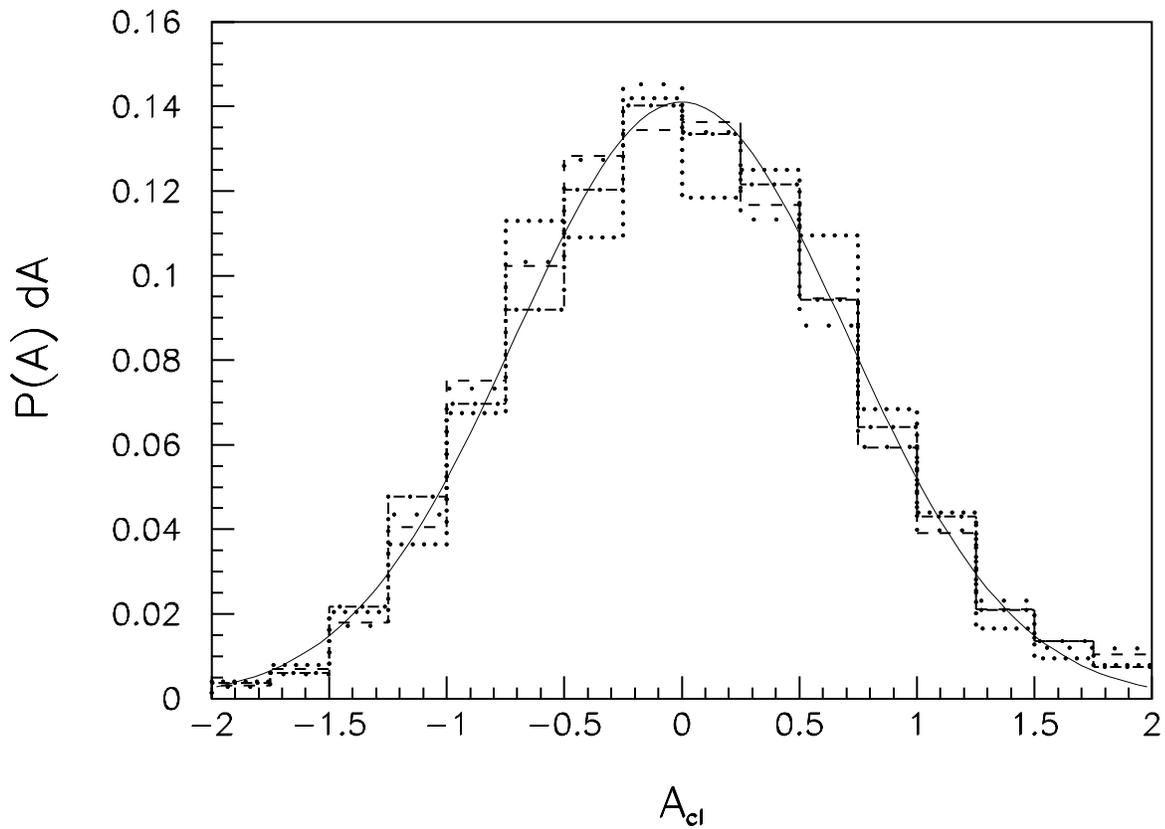

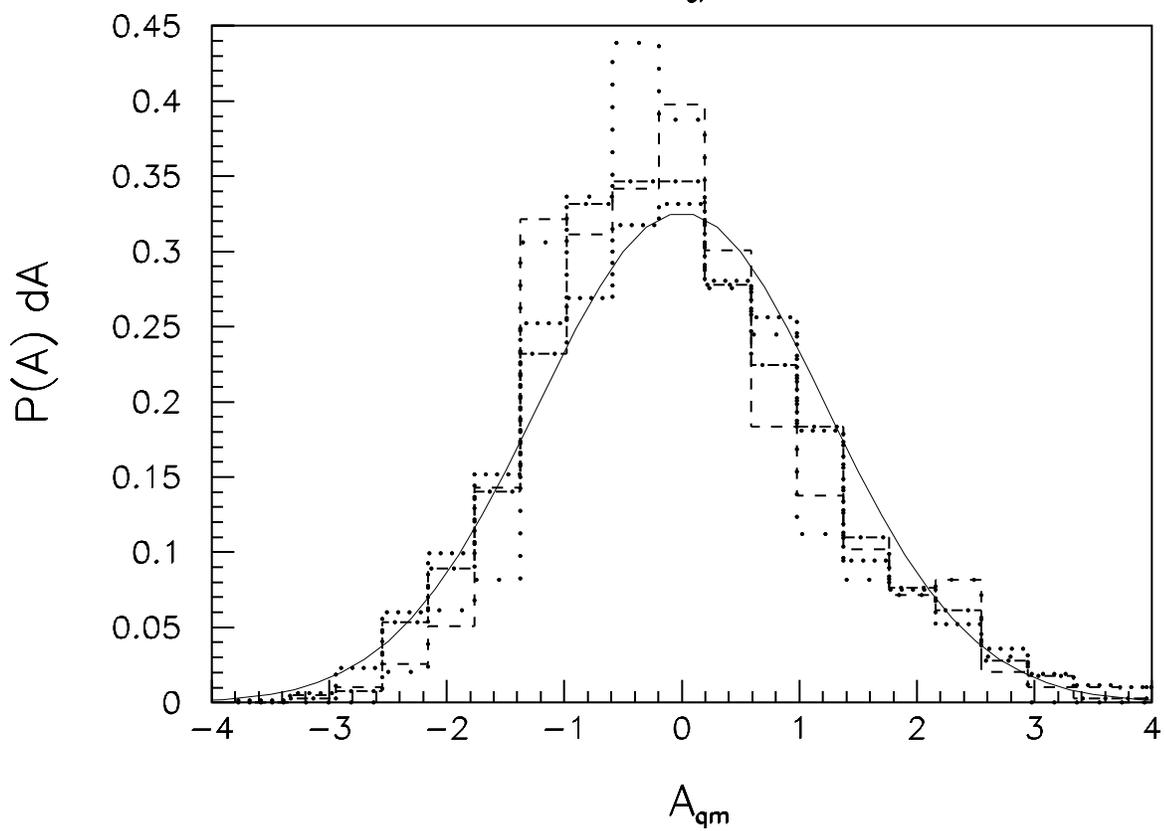

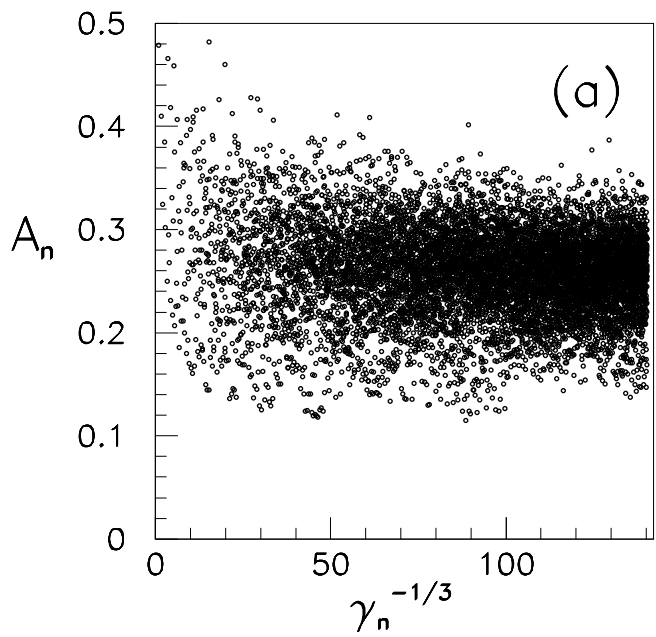 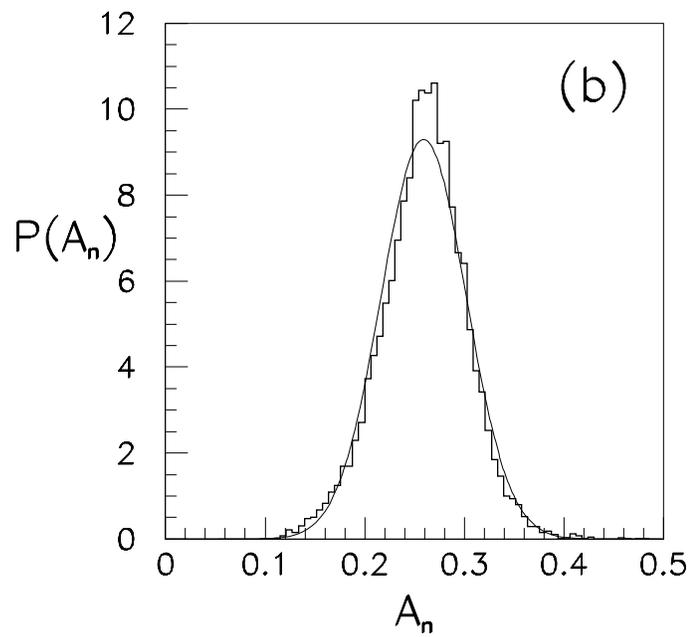

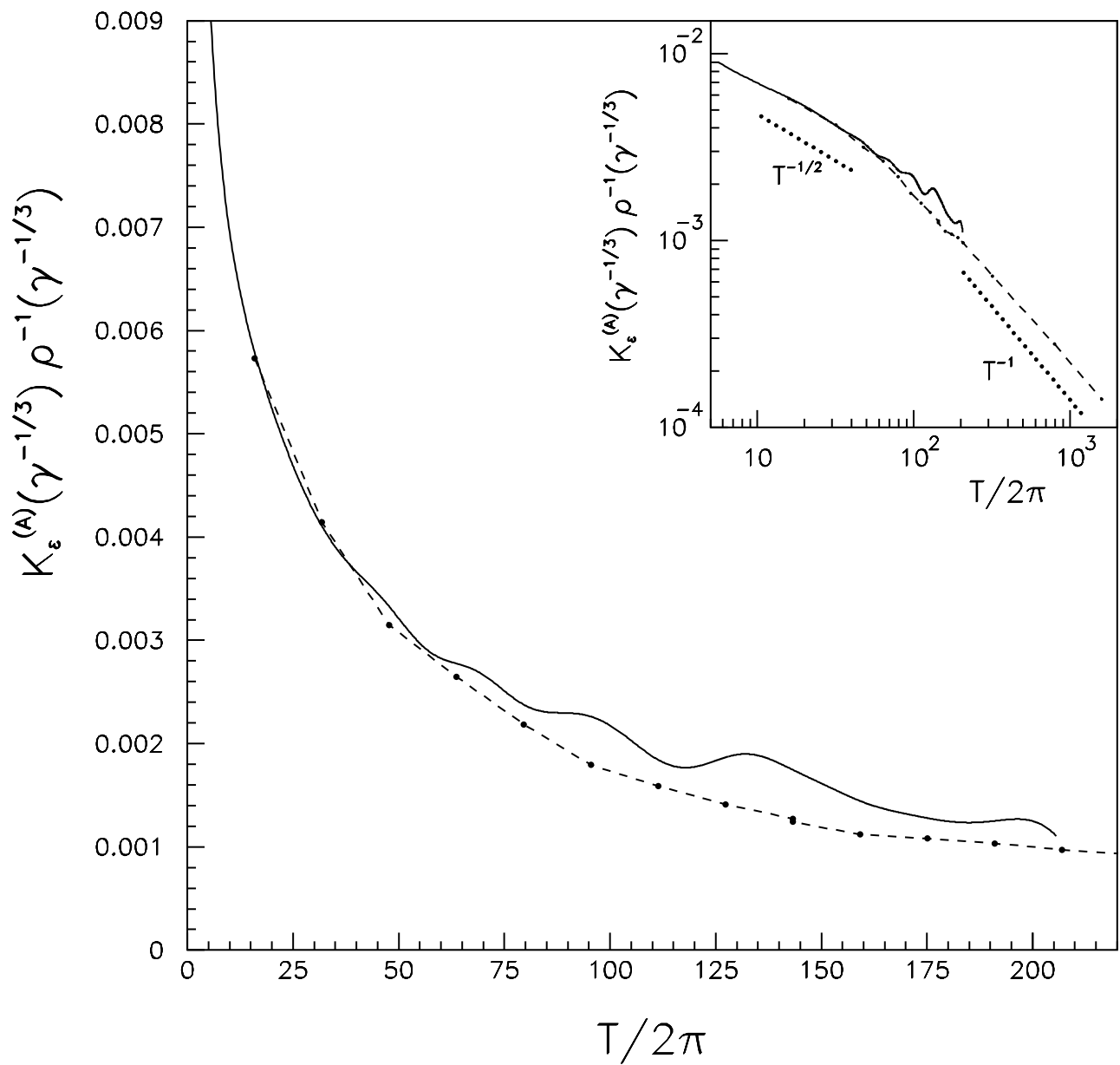